\documentclass[12pt]{article}
\usepackage{amssymb}
\usepackage{epsfig}
\usepackage{float}
\usepackage{slashed}
\newcommand{\be}{\begin{equation}}
\newcommand{\ee}{\end{equation}}
\newcommand{\bea}{\begin{eqnarray}}
\newcommand{\eea}{\end{eqnarray}}

\def\p{\partial}
\def\pslash{\p\raise.3ex \hbox{\kern-.5em /}}
\def\delslash{\nabla\raise.3ex \hbox{\kern-.7em /}}

\begin{document}
\vskip 1cm

\begin{center}
 \textbf{$\cal PT$-Symmetric Pseudo-Hermitian Relativistic Quantum Mechanics With a Maximal Mass }
\end{center}
\vskip 1cm\begin{center} V. N. Rodionov
\end{center}

\begin{center}
{RGU, Moscow, Russia}
\end{center}

\begin{center}
 {\em E-mail    vnrodionov@mtu-net.ru}
\end{center}

\begin{center}

\abstract{The  quantum-field model described by non-Hermitian, but
a ${\cal PT}$-symmetric Hamiltonian is considered. It is shown by
the algebraic way that the limiting  of the physical mass value $m
\leq m_{max}= {m_1}^2/2m_2$ takes place for the case of a fermion
field with a $\gamma_5$-dependent mass term ($m\rightarrow m_1
+\gamma_5 m_2 $). In the regions of unbroken $\cal PT$ symmetry
the Hamiltonian $H$ has another symmetry represented by a linear
operator $ \cal C$. We exactly construct this  operator  by using
a non-perturbative method.  In terms of $ \cal C$ operator we
calculate a time-independent inner product with a positive-defined
norm. As a consequence of finiteness mass spectrum we have  the
$\cal PT$-symmetric Hamiltonian in  the areas $(m\leq m_{max})$,
but beyond this limits $\cal PT$-symmetry is broken. Thus, we
obtain that the basic results of the fermion field model with a
$\gamma_5$-dependent mass term is equivalent to the Model with a
Maximal Mass which for decades has been developed by V.Kadyshevsky
and his colleagues. In their numerous papers the condition of
finiteness of elementary particle mass spectrum was introduced in
a purely geometric way, just as the velocity of light is a maximal
velocity  in the special relativity. The adequate geometrical
realization of the limiting mass hypothesis is added up to the
choice of  (anti) de Sitter momentum space of the constant
curvature. }
\end{center}


\section{Introduction}

In 1965 M. A. Markov  \cite{Markov} pioneered the hypotheses
according to which the mass spectrum of the elementary particles
should be cut off at the Planck mass $m_{Planck} = 10^{19} GeV $ :
\begin{equation}\label{mpl}
 m \leq m_{Planck}.
\end{equation}
The particles with the limiting mass $m = m_{Planck}$, named by
the author "maximons"  should  play special role in the world of
elementary particles. However, Markov's original condition
(\ref{mpl}) was purely phenomenological and he used standard field
theoretical techniques even for describing the maximon.

Till recently one can see no reason why Standard Model (SM) should
not be adequate up to value of order the Planck mass. But we are
living in times, where many of the basic principles of physics are
being challenged by need to go beyond SM. By now it is confirmed
that dark matter exists and it consists of a large fraction of the
energy density of the Universe.

In this connection
 a more radical approach was developed \cite{Kad1} - \cite{Rod}.
The Markov' s idea about existence of a maximal value for the
masses of the elementary particles has been understood as a new
fundamental principle of Nature, which similarly to the
relativistic and quantum postulates should be put in the grounds
of quantum field theory (QFT). Doing this the condition of
finiteness of the mass spectrum should be introduced by the
relation:
\begin{equation}\label{Mfund1}
m \leq  M,
\end{equation}
where the maximal mass parameter M called the "fundamental mass"
is a new universal physical constant.  Now objects for which $ m
> M$ cannot be considered as elementary particles, as  to them does not
correspond a local field.

A new concept of a local quantum field has been developed on the
ground of (\ref{Mfund1}) an on simple geometric arguments, the
corresponding Lagrangians were constructed  and an adequate
formulation of the principle of local gauge invariance has been
found. It has been also demonstrated that the fundamental mass $M$
in the new approach plays the role of an independent universal
scale in the region of ultra high energies $E \geq M $.

The above-presented approach allows a simple geometric realization
if one considers that the fundamental mass $M$ is the curvature
radius of the momentum anti de Sitter 4-space ($\hbar = c =1$)
\begin{equation}\label{O32}
    p_0^2 - p_1^2 - p_2^2 - p_3^2 + p_5 ^2 = M^2.
\end{equation}

For a free particle, for which  $p_0^2 - \overrightarrow{p}^2 =
m^2$, the condition (\ref{Mfund1})  is automatically satisfied on
the surface (\ref{O32}). In the approximation
\begin{equation}\label{Plpred}
    |p_0|,\;\;|\overrightarrow{p}| \ll M, p_5 \cong M.
\end{equation}

\noindent the anti de Sitter geometry does not differ from the
Minkowski geometry in four dimensional pseudo--Euclidean $p$-space
("flat limit").

However, it is much less obvious that in  the momentum 4-space
(\ref{O32}) one may fully develop the apparatus of QFT, which
after transition to configuration representation (with the help of
a specific 5-dimensional Fourier transform) looks like a local
field theoretical formalism in the four dimensional $x$-space
 \cite{Kad1} - \cite{KM}. It is fundamentally important that the new
theory may be formulated in a gauge invariant way \cite{KMNC} -
\cite{KadEch}. In other words, in the considered geometric
approach there are conditions to construct an adequate
generalization of the SM, which was called the Maximal Mass Model
\cite{Max},\cite{KMRS}.

Non-Hermitian quantum mechanics has recently created a lot of
interest.
 This is due to the observation by Bender and Boettcher \cite{ben,ben11}
that with properly defined boundary conditions, the spectrum of
the system described by the Hamiltonian $ H= p^2 + x^2 (ix)^\nu, \
\ \nu \geq 0$ is real positive and discrete. The reality of the
spectrum is a consequence of unbroken $\cal PT$ i.e. combined
Parity $\cal P$ and Time reversal $\cal T$ invariance of the
Hamiltonian, $ [H,PT]\psi =0$ and the spectrum becomes partially
complex when the PT symmetry is broken spontaneously
\cite{pt3,pt31}.

This new result has given rise to growing interest in the
literature, see for examples, \cite{pt3}-\cite{co}. Past few years
many non-Hermitian but $\cal PT$ symmetric systems have been
investigated including  field theoretic models
\cite{ft}-\cite{co}.

In an alternative approach \cite{alir}-\cite{spec1}, it has been
shown that the reality of the spectrum of a non-Hermitian system
is due to so called pseudo-Hermiticity properties of the
Hamiltonian. A Hamiltonian is called $\eta_0 $ pseudo-Hermitian if
it satisfies the relation
\begin{equation}
\eta_0 H \eta_0 ^{-1} = H^\dagger\label{ps},
\end{equation}  where $\eta_0 $ is some
linear Hermitian and invertible operator. All PT symmetric
non-Hermitian systems are pseudo-Hermitian where parity operator
plays the role of $\eta_0 $.

 However,  most of the previous works in the  pseudo-Hermitian quantum
mechanics have been carried out in the non-relativistic framework.
 One of the purpose of this paper is to extend the results of
pseudo-Hermitian quantum mechanics for relativistic systems. Here
we  consider an example of pseudo-Hermitian Hamiltonian and show
that the mass spectrum obtained by solving corresponding Dirac
equation with a $\gamma_5$-mass term is not only real, but should
be cut off $m\leq m_{max} $.

Particularly,  the eigenstates of $\cal PT$ symmetric
non-Hermitian Hamiltonians, with real eigenvalues only, do not
satisfy standard completeness relations.
 More importantly  the eigenstates have negative norms if one
takes the natural inner product associated with such systems
defined as

\begin{equation}
 <f\mid g>_{\cal PT} = \int d^3 x
[{\cal PT} f(x)] g(x).
\end{equation}
 These
problems are overcome by introducing a new symmetry $\cal C$,
analogous to charge conjugation symmetry,  associated with all
such systems with equal number of negative and positive norm
states \cite{pt1,co}. This allows to introduce an inner product
structure associated with $\cal CPT$ conjugate as

\begin{equation}
  <f\mid g>_{\cal CPT} = \int d^3 x [{\cal CPT} f(x)] g(x),
  \label{cpt}
  \end{equation}
for which the norms of the quantum states are positive definite
and one gets usual completeness relation.
 As a result,
the Hamiltonian and its eigenstates can be extended to complex
domain so that the associated eigenvalues are real and underlying
dynamics is unitary. Thus we have a fully consistent quantum
theory for non-Hermitian but $\cal PT$ invariant systems.

 The norms of the state vectors, defined according to the modified
rule of scalar product  will be positive definite
 if we construct the $\cal C$ operator. In Refs \cite{ft12}   it is shown
 that $\cal C$ operator has the general form
\begin{equation}\label{C}
   {\cal C} = e^{Q}P, \label{Q}
\end{equation}
where $Q$ is a Hermitian operator and $P$ is parity operator.
 However, unlike to \cite{ft12}  where  operator $Q$ has been obtained
 perturbatively we can  construct $\cal C$ operator immediately in
 closed form. Similarly the positive definite $\eta $-operator
 in the same way we define.

This paper is arranged as follows. The basic principles of  the
quantum field theory with the  Maximal Mass are considered in
section II. In section III, we formulate  a new algebraic
condition of an unbroken ${\cal PT}$ symmetry. Then the particle
mass finiteness in the theory with a $\gamma_5$ mass term is
obtained and the ${\cal C}$ operator is exactly calculated by
using the non-perturbative method. The positive definite $\eta $
and $\cal C$ -operators we also construct here. Last  IV section
reveals for conclusion and summary

\section{ The Quantum Field Model With a Maximal Mass}

As for the mass of the particle $m$, this quantity is the Casimir
operator of the \emph{{noncompact}}  Poincar\'{e} group and in the
unitary representations of this group, used in QFT, they may have
arbitrary values in the interval $0 \leq m <\infty$. In the SM one
observe a great variety in the mass values. For example, t-quark
is more than 300000 times heavier than the electron. In this
situation  the question naturally arises: up to what values of
mass one may apply the concept of a local quantum field? Formally
the contemporary QFT remains logically perfect scheme and its
mathematical structure does not change at all up to arbitrary
large values of quanta's masses. For instance, the free
Klein-Gordon equation for the one component real scalar field
$\varphi(x)$ has always the form:
\begin{equation}\label{KG}
    (\square + m^2)\varphi(x) = 0.
\end{equation}
From here after standard Fourier transform:
\begin{equation}\label{FT}
    \varphi(x) = \frac{1}{(2\pi)^{3/2}} \int e^{-i p_\mu x^\mu}\;\varphi(p)\;d\,^4
    p\;\;\;\;\; ( p_\mu x^\mu = p^0 x^0 - \emph{\textbf{p.}} \emph{\textbf{x}}
    ),
\end{equation}
we find the equation of motion in Minkowski momentum 4-space:
\begin{equation}\label{KGp}
    ( m^2 - p^2 ) \varphi(p) = 0, \;\;\;\;\;\; p^2 = p_0^2 -
    \emph{\textbf{p}}^2.
\end{equation}
From geometrical point of view $m$ is the radius of the "mass
shell" hyperboloid:
\begin{equation}\label{ms}
    m^2 = p_0^2 -
    \emph{\textbf{p}}^2,
\end{equation}
where the field $\varphi(p)$ is defined and in the Minkowski
momentum space one may embed hyperboloids of the type (\ref{ms})
of arbitrary radius.

It is worth emphasizing that here, due to eq(\ref{Mfund1}), the
Compton wave length of a particle $\lambda_C = \hbar/ m c$ can not
be smaller than the "\emph{{fundamental length}}" $l = \hbar/ M
c$. According to Newton an Wigner \cite {NW} the parameter
$\lambda_C$ characterizes the dimensions of the region of space in
which a relativistic particle of mass $m$ can be localized.
Therefore the fundamental length $l$ introduces into the theory an
universal bound on the accuracy of the localization in space of
elementary particles.

Let us go back to the free one component real scalar field we
considered above (\ref {KG} - \ref{KGp}). We shall suppose that
its mass $m$ satisfies the condition (\ref{Mfund1}). How  should
one modify the equations of motion in order that the existence of
the bound (\ref{Mfund1}) should become as evident as it is the
limitation $v \leq c$ in the special theory of relativity? In the
latter case everything is explained in a simple way: the
relativization of the 3-dimensional velocity space is equivalent
to transition in this space from Euclidean to Lobachevsky
geometry, realized on the 4-dimensional hyperboloid.  Let us act
in  a similar way and change the 4-dimensional Minkowski momentum
space, which is used in the standard QFT, to anti de Sitter
momentum space, realized on the 5-hyperboloid:
\begin{equation}\label{ads}
    p^2_0 - \emph{\textbf{p}}^2 + p_5^2 = M^2.
\end{equation}

\begin{figure}[h]
\vspace{-0.2cm} \centering
\includegraphics[angle=0, scale=0.5]{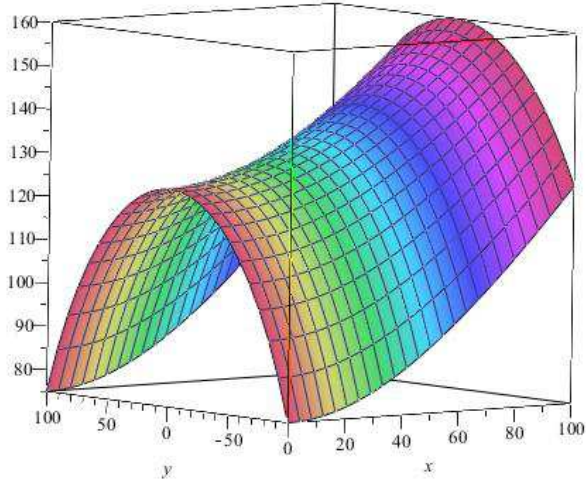}
\caption{Curvature momentum space, realized on  the hyperboloid
${p_0}^2 -{p_1}^2 + {p_5}^2 =M^2$, for $M = 125 GeV$.}
\vspace{-0.1cm}\label{Fig.1}
\end{figure}
In Fig.~\ref{Fig.1} we have the 3D-plot of $P_0$ as function of
$P_1$ and $P_5$,
 for the case $P_2=P_3$=0 and maximal mass $M=125 GeV$.

We shall suppose that in p-representation our scalar field is
defined just on the surface (\ref{ads}), i.e. it is a function of
five variables $(p_0, \emph{\textbf{p}}, p_5)$, which are
connected by the relation (\ref{ads}):
\begin{equation}\label{dads}
    \delta(p^2_0 - \emph{\textbf{p}}^2 + p_5^2 - M^2)\varphi(p_0, \emph{\textbf{p}},
    p_5).
\end{equation}
The energy $p_0$ and the 3-momentum $\emph{\textbf{p}}$ here
preserve their usual sense and the mass shell relation (\ref{ms})
is satisfied as well. Therefore,  for the considered field
$\varphi(p_0, \emph{\textbf{p}}, p_5)$ the condition
(\ref{Mfund1}) is always fulfilled.

Clearly in eq. (\ref{dads}) the specification of a single function
$\varphi(p_0, \emph{\textbf{p}}, p_5)$ of five variables $(p_\mu,
p_5)$ is equivalent to the definition of two independent functions
$\varphi_1(p)$ and $\varphi_2(p)$ of the 4-momentum $p_\mu$:
\begin{equation}\label{12}
    \varphi(p_0, \emph{\textbf{p}}, p_5) \equiv \varphi(p, p_5) = \left(%
\begin{array}{c}
  \varphi(p, p_5) \\
  \varphi(p, - p_5) \\
\end{array}%
\right) = \left(%
\begin{array}{c}
  \varphi_1(p) \\
  \varphi_2(p) \\
\end{array}%
\right), |p_5| = \sqrt{M^2 - p^2}.
\end{equation}

The appearance of the new discrete degree of freedom $p_5/|p_5|$
and the associated doubling of the number of field variables is a
most important feature of the new approach. It must be taken into
account in the search of the equation of motion for the free field
in de Sitter momentum space. Because of the mass shell relation
(\ref{ms}) the Klein - Gordon equation (\ref{KGp}) should be also
satisfied by the field $\varphi(p_0, \emph{\textbf{p}}, p_5)$ :
\begin{equation}\label{KGp2}
    ( m^2 - p_0^2 + \emph{\textbf{p}}^2)  \varphi(p_0, \emph{\textbf{p}}, p_5) =
    0.
\end{equation}
From our point of view this relation is unsatisfactory for 2
reasons:

1. It does not reflect the bounded mass condition (\ref{Mfund1}).

2. It can not be used to determine the dependence of the  field on
the new quantum number $p_5/|p_5|$ in order to distinguish between
the components $\varphi_1(p)$ and  $\varphi_2(p)$.

Here we notice that, because of (\ref{ads}) eq.(\ref{KGp2}) may be
written as:
\begin{equation}\label{razlKG}
    (p_5 + M \cos \mu)(p_5 - M \cos \mu)\varphi(p, p_5) = 0 , \;\;\;\;\; \cos
    \mu = \sqrt{1 - \frac{m^2}{M^2}}.
\end{equation}
Now, following the  Dirac trick we postulate the  equation of
motion under question in the form:
\begin{equation}\label{NKG}
    2M(p_5 - M \cos \mu)\varphi(p, p_5) = 0 .
\end{equation}
Clearly, eq. (\ref{NKG}) has none of the enumerated defects
present in the standard Klein-Gordon equation (\ref{KGp}).
However, equation (\ref{KGp}) is still  satisfied by the field
$\varphi(p, p_5)$.

From eqs. (\ref{NKG}) and (\ref{12}) it follows that:
\begin{equation}\label{NKG12}
    \begin{array}{c}
      2M(|p_5| - M \cos \mu)\varphi_1(p) = 0, \\ \\
      2M(|p_5| + M \cos \mu)\varphi_2(p) = 0, \\
    \end{array}
\end{equation}
and we obtain:
\begin{equation}\label{122}
    \begin{array}{c}
      \varphi_1(p) = \delta(p^2 - m^2)\widetilde{\varphi}_1(p)\\ \\
      \varphi_2(p) = 0.\\
    \end{array}
\end{equation}
Therefore, the free field $\varphi(p, p_5)$ defined in anti de
Sitter momentum space (\ref{ads}) describes the same free scalar
particles of mass $m$ as the field $\varphi(p)$ in Minkowski
p-space, with the only difference that now we necessarily have $m
\leq M$. The two component structure (\ref{12}) of the new field
does not manifest itself on the mass shell, owing to (\ref{122}).
However, it will play an important role when the fields interact
- i.e off the mass shell.

In the ordinary formalism the free Dirac operator \be\label{K1}
D(p) = p_\nu \gamma^\nu - m ;
 \nu=0,1,2,3
 \ee
  appears as a result of
factorization of the Klein-Gordon wave operator \be\label{K2}
p_\nu^2 - m^2 = (p_\nu\gamma^\nu + m)(p_\nu\gamma^\nu -m ). \ee
Now instead of (\ref{K1}),(\ref{K2}) we obtain the following
factorization formulas:

\begin{equation}\label{1}
2 M (p_5+M {\rm cos}\mu)=\big[\gamma^0 p_0 - {\bf \gamma p}-
\end{equation}
$$
\gamma^5(p_5+M)-2 M {\rm sin}\mu/2\big ] \big[\gamma^0
p_0-{\bf\gamma p}-\gamma^5(p_5+M)+2 M {\rm sin}\mu/2\big ].
$$

\begin{equation}\label{2}
2 M (p_5-M{\rm cos}\mu)=\big[\gamma^0 p_0-{\bf{\gamma p}}
\end{equation}
$$
-\gamma^5(p_5-M)+ 2 M {\rm sin}\mu/2\big]\big[-\gamma^0
p_0+{\bf\gamma p}+\gamma^5(p_5-M)+2M {\rm sin}\mu/2\big].
$$
There is instead of (\ref{K1}) we have the new expression for the
Dirac operator

\be \label{K3} D(p,M)= p_\nu \gamma^\nu + (p_5-M)\gamma^5
+2M\sin(\mu/2). \ee
 It is easy to check that in the "flat
approximation"
$$
|p_\nu|\ll M,\;\;\;m\ll M, p_5\ll M, p_5\simeq M.
$$
both expressions (\ref{K1}),(\ref{K3}) coincide. But the amusing
point is that the new Klein-Gordon operator $2M(p_5- \cos\mu)$ has
one more decomposition into matrix factors:

 \be \label{K4} 2M(p_5- \cos\mu)= \big[\gamma^0 p_0-{\bf{\gamma p}}
\end{equation}
$$
-\gamma^5(p_5+M)+ 2 M {\cos}\mu/2\big]\big[-\gamma^0
p_0+-{\bf\gamma p}+\gamma^5(p_5+M)-2M {\cos}\mu/2\big].
$$
Therefore, if our approach is considered to be realistic, it may
be assumed that in Nature there exists some exotic fermion field
associated with the wave operator

\be \label{K5} D_{exotic}(p,M)= p_\nu \gamma^\nu + (p_5+M)\gamma^5
-2M\cos(\mu/2). \ee In contrast  to (\ref{K3}) the operator
$D_{exotic}(p,M) $ dose not have a flat limit
$(M\longrightarrow\infty)$.

\section{ The Particle Mass Finiteness in The Theory With a $\gamma_5$ - Mass
Term}

Now we can consider (\ref{1}),(\ref{2})in configuration space  on
the mass surface $p_5 = \pm\sqrt{M^2 - m^2} $ (\cite{Max}).

For the case $p_5=- \sqrt{M^2-m^2}$ we have
\begin{equation}\label{3}
\Big(p_0 - {\bf \alpha p}- \beta m_1 - \beta\gamma^5 m_2\Big
)\Psi_1(x,t,x_5)=0.
\end{equation}
$$
\Big(p_0 - {\bf \alpha p}+ \beta m_1 - \beta\gamma^5 m_2\Big
)\Psi_2(x,t,x_5)=0.
$$
Analogously, for the case $p_5= \sqrt{M^2-m^2}$ we can write
\begin{equation}\label{4}
\Big(p_0 - {\bf \alpha p}- \beta m_1 + \beta\gamma^5 m_2\Big
)\Psi_3(x,t,x_5)=0.
\end{equation}
$$
\Big(p_0 - {\bf \alpha p}+ \beta m_1 + \beta\gamma^5 m_2\Big
)\Psi_4(x,t,x_5)=0.
$$
In this equations Dirac matrices $\beta =\gamma_0$,
$\gamma^i=\beta\alpha^i$.

Equation (\ref{3}),(\ref{4}) differ from each other only in their
signs before the terms with $m_1$ and $m_2$. As for their physical
this equation are equivalent to the ordinary Dirac equations
differing in their signs before the mass term. It is very
important to note, that on the mass surface there are not the
operators, which act on the coordinate $x_5$ and this parameter
can be taken equal to zero (\cite{Max}),(\cite{KMRS}).

Any of the equivalent Hamiltonian from
equations(\ref{1}),(\ref{2}) takes the form

 \be\label{H1} {\hat{H}} =
\overrightarrow{\hat{\alpha}}\overrightarrow{p} + \hat{\beta
}\left(m_1 +m_2\gamma_5\right) \ee We consider now the case of
two-dimensional space-time. In $(1+1)$-dimensional space-time we
adopt the conventions used in Ref.~\cite{2D}: \be
\gamma_0=\left(\begin{array}{cc}0 & 1\\ 1 & 0
\end{array}\right) \quad{\rm and}\quad
\gamma_1=\left(\begin{array}{cc}0 & 1\\ -1 & 0 \end{array}\right).
\label{e14} \ee
 With these definitions we have $\gamma_0^2=1$ and
$\gamma_1^2=-1$. We also define \be \gamma_5=
-\gamma_0\gamma_1=\left(\begin{array}{cc}1&0\\0&-1\end{array}\right),
\label{e15} \ee
 so that $\gamma_5^2=1$.

It is easy to see that the Hamiltonian $H_0=
\overrightarrow{\hat{\alpha}}\overrightarrow{p} + \hat{\beta
}\left(m\right)$ is Hermitian: $H_0=H_0^\dagger$. Also, $H_0$ is
separately invariant under parity reflection and under time
reversal:
$${\cal P}H_0{\cal P}=H_0\quad{\rm and}\quad {\cal T}H_0{\cal T}=H_0.$$

Now let us consider a {\it non}-Hermitian Hamiltonian
(\ref{H1}).The Hamiltonian $H$  is not Hermitian because the $m_2$
term changes sign under Hermitian conjugation.  Also, $H$ is not
invariant under $\cal P$ or under $\cal T$ separately because the
$m_2$ term changes sign under each of these reflections. This sign
change occurs because $\hat\beta$ and $\gamma_5$ anticommute.
However, $H$ {\it is invariant} under combined $\cal P$ and $\cal
T$ reflection. Thus, $H$ is $\cal PT$-symmetric: $H^{\cal
PT}={\cal PT}H{\cal PT}=H.$

The equation (\ref{4}) can be transformed to the equation of the
second order and a result we have
 the Klein-Gordon equation
 \be \label{KG}
\left(\partial^2+m^2\right)\psi(x,t)=0, \label{e20}
 \ee
  where
\be\label{m} m^2=m_1^2-m_2^2. \ee

It easy to see the physical mass $m$ that propagates under this
equation is real when the inequality \be\label{m1m2} m_1^2\ge
m_2^2 \label{e21} \ee is satisfied. This inequality (\ref{m1m2})
was considered by C.Bender et al. in Ref.\cite{ft12} as the basic
condition.


\begin{figure}[h]
\vspace{-0.2cm} \centering
\includegraphics[angle=0, scale=0.5]{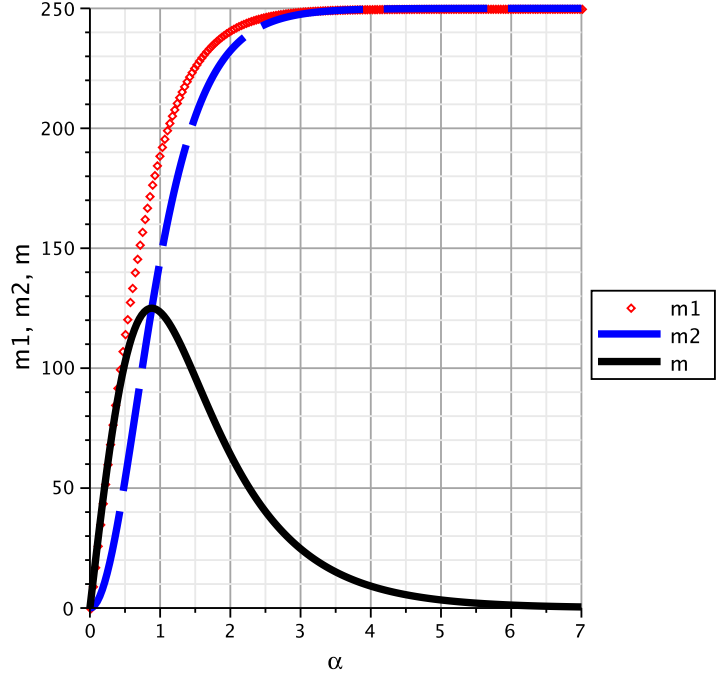}
\caption{The values of parameters m, m1,m2 as  functions of
$\alpha$; m(max)=125 GeV.} \vspace{-0.1cm} \label{f3}
\end{figure}

Now we prove that the parameters $ m_1$ and $ m_2$   have
auxiliary  nature because that  assume  an ambiguous definition.
 Taking into account (\ref{m}), we can write the
following obvious inequality \be
 (m-m_2)^2 \geq 0,
\ee from which to obtain
 \be \label{M}m \leq \frac{{m_1}^2}{2
m_2}= m_{max}.\ee The  conditions (\ref{m}),(\ref{e21}) and
(\ref{M}) are satisfied automatically  if we introduce  the
following parametrization: \be\label{alpha00} m_1 = m
\cosh(\alpha);\;\;\;\;\ m_2= m \sinh(\alpha). \ee Indeed, from
(\ref{M}) and (\ref{alpha00}) we can also define

\be\label{alpha0} m = 2 m_{max}\frac{\sinh\alpha}{\cosh^2\alpha},
\ee

\be\label{alpha1} m_1 = 2 m_{max}\tanh\alpha, \ee

\be\label{alpha2} m_2 = 2 m_{max}\tanh^2\alpha.\ee
  For this mass  the all conditions
(\ref{m}), (\ref{e21}) and (\ref{M}) are realized. Fig.~(\ref{f3})
displays parameters $m$, $m_1$ and $m_2$ as functions of $\alpha$
are presented. Value entry of the maximal mass is $125 GeV$. The
values of parameters $m$, $m_1$,$m_2$  describes the propagation
of particle having positive mass $m$ can be varied in a wide
range. The $m$ reaches a high ($m=m_{max} $) in the point
$\alpha_0 = 0.881 $.

The equation (\ref{M}) can be represented in the form
 \be \label{tan}
\frac{m}{2 m_{max}} = \tanh(\alpha)\sqrt{1-\tanh^2(\alpha)}. \ee
The solution of (\ref{tan}) is

\be\label{tan1} \tanh(\alpha) = \sqrt{\frac{1
\pm\sqrt{1-m^2/m^2_{max}}}{2}}. \ee Two signs of square root can
be interpreted by the following way: we have the dual branches of
the values $m_1$ and $m_2$ as functions of the physical mass $m$.

This is the reason to insert a new definitions for mass.  Using
(\ref{alpha1}),(\ref{alpha2}) we have

\be\label{m1} m_1 = \sqrt{2} m_{max}\sqrt{1
-\sqrt{1-m^2/m^2_{max}}};
 \ee

\be \label{m2}m_2 =m_{max}\left( 1 -\sqrt{1-m^2/m^2_{max}}\right);
\ee

 \be\label{m3} m_3 = \sqrt{2} m_{max}\sqrt{1 +\sqrt{1-m^2/m^2_{max}}} \ee

\be\label{m4} m_4 ==m_{max}\left( 1
+\sqrt{1-m^2/m^2_{max}}\right). \ee

\begin{figure}[h]
\vspace{-0.2cm} \centering
\includegraphics[angle=0, scale=0.5]{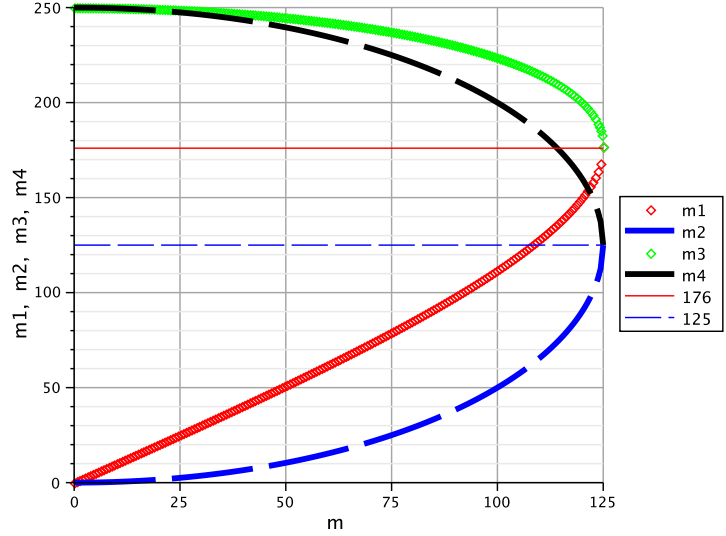}
\caption{The values of parameters $m1, m2, m3, m4$ as  functions
of $m$;  $M=125 GeV.$} \vspace{-0.1cm}\label{f4}
\end{figure}

Fig.~(\ref{f4}) displays  parameters $m_1$, $m_2$,$m_3$ and $m_4$
as functions of $m$ values are presented. The region of unbroken
$\cal PT$ symmetry $m\leq m_{max}$. For these values of the
parameters $m_1$ and $m_2$, the  new Dirac equation describes the
propagation of particles having the real mass. The special case of
Hermiticity is obtained on the line $m=m_{max}$, which is achieved
at the edge of the region of unbroken $\cal PT$ symmetry. In this
point we have $m_1=m_3=\sqrt{2}m_{max}$ and $m_2=m_4=m_{max}$.

There is no difficulty in understanding that a new mass parameters
are really  satisfied to conditions (\ref{m}), (\ref{e21}). In
particular, we have $m_3\geq m_4$ and

\be {m_3}^2 -{m_4}^2 =m^2. \ee The condition (\ref{M}) and an
ambiguous definition of $m_1$,$m_2$
 are in agreement with
the Kadyshevsky's basic principal of the geometrical scheme
\cite{Max},\cite{KMRS}.



This approach may be used for the calculation  of the $\cal C$
operator associated with  the $\cal PT$-symmetry of Hamiltonian.
We begin by letting (\ref{alpha00}) and rewriting the mass terms
in Hamiltonian in the form

\be \hat\beta(m_1 +
m_2\gamma_5)=\hat{\beta}m(\cosh\alpha+\gamma_5\sinh\alpha)=\hat{\beta}m\exp{(\gamma_5\alpha)}.
\ee Then, Hamiltonian $\hat H$ is given by

\be \label{H} \hat H= \hat{\overrightarrow{\alpha}}
\overrightarrow{p}+\hat{\beta}m\exp({\gamma_5\alpha}), \ee and
Hermitian-conjugate Hamiltonian takes the form \be \hat{H^+}=
\hat{\overrightarrow{\alpha}}
\overrightarrow{p}+\hat{\beta}m\exp({-\gamma_5\alpha}). \ee Next,
we can write

\be\label{H}
    e^{\hat{\alpha}\gamma_5 /2} {\hat{H}} =\left(
     \overrightarrow{\hat{\alpha}}\overrightarrow{p} + \hat{\beta }m\right)e^{\hat{\alpha}\gamma_5
    /2}={\hat{H_0}}e^{\hat{\alpha}\gamma_5 /2},
\ee where  the sign before the mass term $m_2$ change occurs
because the $\gamma_5$ and $\beta$ anticommute and
$$\hat H_0 =
\overrightarrow{\hat{\alpha}}\overrightarrow{p} + \hat{\beta }m
$$
is the ordinary Dirac Hamiltonian.

For Hamiltonian $\hat{H^+}$ we also have

\be\label{H+}
    e^{-\hat{\alpha}\gamma_5 /2} {\hat{H^{+}}} =\left(
    \overrightarrow{\hat{\alpha}}\overrightarrow{p} + \hat{\beta }m\right)e^{-\hat{\alpha}\gamma_5
    /2}={\hat{H_0}}e^{-\hat{\alpha}\gamma_5 /2}.
\ee
 It is easy to see from (\ref{H}),(\ref{H+}) Hermitian
Hamiltonian $\hat H_0$ ($\hat H_0 = \hat {H_0}^+$) and $\hat
H$,$\hat H^+$ are related by similarity transformations

\be \label{H0} \hat {H_0} = e^{\frac{\gamma_5\alpha}{2}}\hat {H}
e^{\frac{-\gamma_5\alpha}{2}}. \ee

\be \label{H0+} \hat {H_0} = e^{-\frac{\gamma_5\alpha}{2}}\hat
{H^+} e^{\frac{\gamma_5\alpha}{2}}. \ee

From (\ref{H0}),(\ref{H0+}) we have
      \be
e^{-Q}{\hat H}e^{Q} = {\hat H^+}, \ee where
$$
Q  = -\alpha\gamma_5.
$$
In Refs {\cite{ft}},{\cite{ft12}}  it is shown that the ${\cal C}$
operator has the general form (\ref{C}) and in the  case of
(1+1)-dimensional space-time, ${\cal C}$ operator for the model
with $\gamma_5$-mass term can be presented as \be \label{C1}{\cal
C}=\left(\begin{array}{cc}0& \frac{m_1-m_2}{m}\\\frac{m_1 +
m_2}{m}&0\end{array}\right). \ee One make sure that the operator
$\cal C$  is satisfied to the following system of three algebraic
conditions: \bea
{\cal C}^2&=&1, \label{e2}\\
\left[{\cal C},{\cal PT}\right]&=&0, \label{e3}\\
\left[{\cal C},H\right]&=&0.  \label{e4} \eea By solving these
three simultaneous equations for the operator $\cal C$, one
obtains an inner product with respect to which $H$ is
self-adjoint.

In Ref.\cite{ali} it is shown that the square root of the positive
operator
$$
\eta\equiv \sqrt{e^{-Q}}
$$
can be used to contract a Hermitian Hamiltonian $\hat{H_0}$ that
corresponds to the non-Hermitian Hamiltonian $ \hat{H}$.
From(\ref{H0}),(\ref{H0+}) we  can obtain
\begin{equation}\label{eta}
\eta = e^{\alpha\gamma_5 /2}.
\end{equation}

Now we construct  the  norm of any state for considered model
using  $\cal CPT$-symmetry. For arbitrary vector
$$
\Psi= \left( \begin{array}{cc}
 x+i y&{} \\
u+iv&{}
\end{array}\right),
$$
we have

$$
{\cal CPT} \Psi=\left( \frac{m_1-m_2}{m}(x-iy),
\frac{m_1+m_2}{m}(u-iv)\right).
$$
Then \be \label{Psi} \langle{\cal C P T}
\Psi|\Psi\rangle=\frac{m_1-m_2}{m}(x^2+y^2)+\frac{m_1+m_2}{m}(u^2+v^2),
\ee is explicitly non negative, because  $m_1\geq m_2$.

\section{Conclusion }

The investigations given in the previous sections show that the
Dirac Hamiltonian of a particle with the a $\gamma_5$-dependent
mass term ($m\rightarrow m_1 +\gamma_5 m_2 $) is non-Hermitian but
a $\cal PT$ symmetric. It is shown by the algebraic way that the
limiting of the physical mass value $ m_{max}= {m_1}^2/2m_2$ takes
place. In the regions of unbroken ($m\leq m_{max}$) $ \cal PT$
symmetry the Hamiltonian $H$ has another symmetry represented by a
linear operator $ \cal C$ (\ref{C}).

\begin{figure}[h]
\vspace{-0.2cm} \centering
\includegraphics[angle=0, scale=0.5]{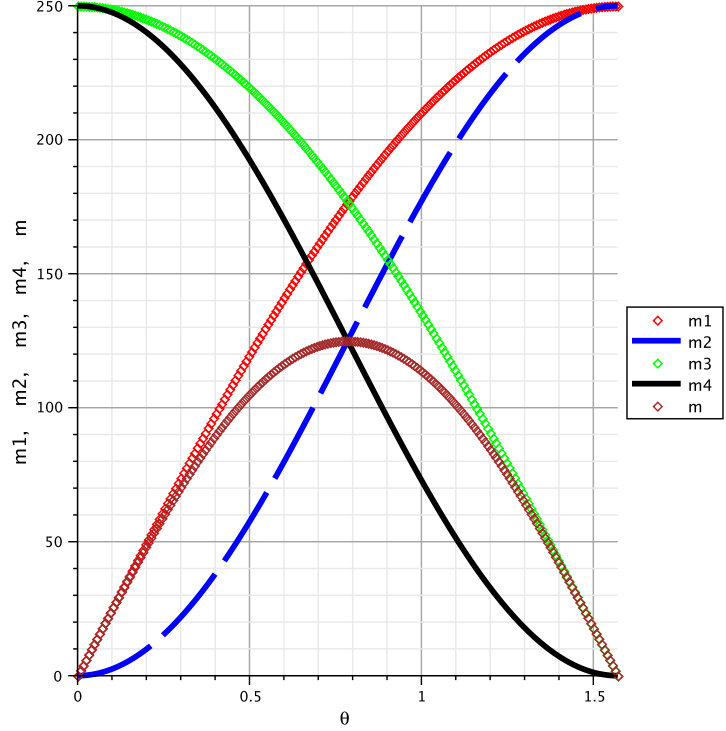}
\caption{The values of parameters $m1, m2, m3, m4, m$ as functions
of $\theta$;  $M=125 GeV.$} \vspace{-0.1cm}\label{f5}
\end{figure}

We exactly construct this  operator  (\ref{C1}) by using a
non-perturbative method.  In terms of $ \cal C$ operator we
calculate a time-independent inner product with a positive-defined
norm. As a consequence of finiteness mass spectrum we have  the
$\cal PT$-symmetric Hamiltonian in  the areas $m\leq m_{max}$, but
beyond this limits $\cal PT$-symmetry is broken.

We proved that the parameters $ m_1$ and $ m_2$   have auxiliary
 nature because  assume  an ambiguous definition. This fact can be
 also
 confirmed by  making a comparison between ordinary(having a flat limit) and "exotic
 fermion field" which dose not have a limit when $m_{max} \rightarrow\infty $.
  Let write a new definitions of
 mass $ m_3$ and $ m_4$ for the exotic  field to satisfy
 conditions

 \be \label{ex}{m_3}^2=m^2+{m_4}^2
 \ee
  $m=m_3 \sin\theta;\;\;m_4=m_3\cos\theta.
 $   Then we have
 $$
m_{max}=\frac{m}{\sin 2\theta};\;\;  m_3 = m_{max}\cos\theta;
\;\;\;\; m_4 = m_{max}\cos^2\theta;\;\;\;   0 \leq \theta \leq
\pi/2.
$$

Analogously,  for ordinary field (${m_1}^2=m^2+{m_2}^2$) we have
$m=m_1\cos\theta;\;\;$ $m_4=m_1\sin\theta  $; and can obtain
$$
m_{max}=\frac{m}{\sin 2\theta};\;\;  m_1 = m_{max}\sin\theta;
\;\;\;\; m_2 =m_{max}\sin^2\theta;\;\;\; 0 \leq \theta \leq \pi/2.
$$

Fig.~(\ref{f5}) displays  parameters $m_1$, $m_2$,$m_3$ and $m_4$
as functions of $\theta$ values are presented. The region of
unbroken $\cal PT$ symmetry  $m\leq m_{max}$ and in term of the
parameter $ \theta $ one can write as $$ 0 \leq \theta \leq \pi/2.
$$
 For these values of the parameters $m_1$ and $m_2$  the new
Dirac equation describes the propagation of particles having the
real mass. The special case of Hermiticity is obtained on the line
$m=m_{max}$, which is achieved at the center of the region of
unbroken $\cal PT$ symmetry $\theta_0=\pi/4 $. In this point we
have $m_1=m_3=\sqrt{2}m_{max}$ and $m_2=m_4=m_{max}$.

 Thus, we obtain
that the basic results of the fermion field model with a
$\gamma_5$-dependent mass term is equivalent to the Model with a
Maximal Mass which for decades has been developed by V.Kadyshevsky
and his colleagues \cite{Kad1} - \cite{Rod}. In particular, the
exotic fermion field (\ref{K5}) associated with the new wave
operator which does not have a limit when $ M
\longrightarrow\infty$ was investigated. The polarization
properties of such the exotic fermion field differ sharply from
the standard ones. It is tempting to think that quanta of the
exotic fermion field have a relation to the structure of the "dark
matter".

{\bf Acknowledgment:} We are grateful to Prof. V.G.Kadyshevsky for
fruitful and highly useful discussions.

\end{document}